\documentclass[twocolumn,showpacs,preprintnumbers,amsmath,amssymb,prl,aps]{revtex4}
\usepackage{graphicx}
\usepackage{natbib}
\usepackage{color} 

\newcommand{\etal}{\emph{et al.}}                    
\newcommand{\bean}{\begin{eqnarray}}                
\newcommand{\eean}{\end{eqnarray}}                  
\newcommand{\bea}{\begin{eqnarray*}}                
\newcommand{\eea}{\end{eqnarray*}}                  
\newcommand{\half}[1][1]{{\textstyle\frac{#1}{2}}}  

\usepackage{multirow}
\usepackage{amsmath}
\usepackage{amssymb}
\usepackage{natbib} 
\usepackage{dcolumn}

\graphicspath{}

\begin{document}

\title{Electrical control of Kondo effect and superconducting transport in a side-gated InAs quantum dot Josephson junction\\}

\author{Y. Kanai$^{1}$}\email{kanai@meso.t.u-tokyo.ac.jp}
\author{R.S. Deacon$^{1}$}\email{russell@meso.t.u-tokyo.ac.jp} 
\author{A. Oiwa$^{1,2,3}$}
\author{K. Yoshida$^{2}$}
\author{K. Shibata$^{4}$}
\author{K. Hirakawa$^{3,4,5}$}
\author{S. Tarucha$^{1,2,5}$}
\affiliation{$^{1}$Department of Applied Physics and QPEC, The University of Tokyo, 7-3-1 Hongo, Bunkyo-ku, 113-8656, Japan.}
\affiliation{$^{2}$ Quantum Spin Information Project, ICORP, JST, Atsugi-shi, Kanagawa 243-0198, Japan.}
\affiliation{$^{3}$ JST CREST, 4-1-8 Hon-cho, Kawaguchi-shi, Saitama 332-0012, Japan.}
\affiliation{$^{4}$Institute of Industrial Science, The University of Tokyo, 4-6-1 Komaba, Meguro-ku, Tokyo 153-8505, Japan.}
\affiliation{$^{5}$ INQIE, The University of Tokyo, 4-6-1 Komaba, Meguro-ku, Tokyo 153-8505, Japan.}

\date{\today}
\begin{abstract}
We measure the non-dissipative supercurrent in a single InAs self-assembled quantum dot (QD) coupled to superconducting leads. The QD occupation is both tuned by a back-gate electrode and lateral side-gate. The geometry of the side-gate allows tuning of the QD-lead tunnel coupling in a region of constant electron number with appropriate orbital state. Using the side-gate effect we study the competition between Kondo correlations and superconducting pairing on the QD, observing a decrease in the supercurrent when the Kondo temperature is reduced below the superconducting energy gap  in qualitative agreement with theoretical predictions.
\end{abstract}

                             
\maketitle

Devices which combine the gate tuneability of semiconductor quantum dots (QDs) with non-dissipative superconducting transport are desirable for very sensitive and controllable coherent switching devices\cite{Paper:VanDam-NAT-2006,Paper:Jarillo-Herrero-NAT-2006} and the study of interplay between Kondo physics and the superconducting proximity effect\cite{Paper:Yeyati-Rodero-2003,Paper:Choi-PRB-2004,Paper:Siano-PRL-2004,Paper:Karrasch-PRB-2008,Paper:Oguri-JPSJ-1994,Paper:Matveev-JETP-1989}. The important energy scales for the interaction between Kondo singlet state and superconductivity are captured in the scaling parameter $t_{K}=k_{B}T_{K}/\Delta$ determined by the superconducting gap ($\Delta$) and Kondo temperature ($T_{K}$). For $t_{K}\gg 1$ the local magnetic moment of the unpaired electron spin on the QD is screened by the Kondo effect and the ground state of the system is a Kondo singlet state. In this regime an enhanced supercurrent due to the Kondo effect has been predicted theoretically\cite{Paper:Matveev-JETP-1989}. However, for $t_{K}\ll 1$ the Kondo state is suppressed by the lack of low energy excitations in the superconducting energy gap and the system ground state is a degenerate (so called magnetic) doublet state. To date experimental efforts to elucidate this phase transition have been unable to systematically control the physical parameters which determine $t_{K}$ due to the limited tuneability of the devices studied. In the present paper, we demonstrate that $t_{K}$ may be smoothly controlled by a side-gate, which is placed laterally to an InAs self-assembled QD. The sidegate performance is very effective for uncapped InAs self-assembled QDs as the lateral confinement tuned by the sidegate is weak relative to that in nanowire or nanotube devices. When $t_{K}$ is tuned through  $t_{K}\sim 1$ we observe a dramatic change in the superconducting transport indicating the phase transition between Kondo singlet and magnetic doublet states.

To date works on interplay of Kondo and proximity effect have focused on either the dissipative transport or the non-dissipative supercurrent. For low biases red{within the subgap transport region the dissipative transport current is carried by multiple Andreev reflections\cite{Paper:Schonenberger-PRL-2003} (MAR). MAR resonances occur when a sequence of Andreev reflections connect the high density of states at the edge of the superconducting energy gap. Sub-gap transport MAR resonances have been shown to be substantially altered by the single electron states of a QD\cite{Paper:Yeyati-PRB-1997, Paper:Johansson-PRB-1999, Paper:Schonenberger-PRL-2003} and the Kondo effect\cite{Paper:Lindelof-PRL-2007, Paper:Schonenberger-PRL-2007,Paper:Buizert1}. Observation of the supercurrent in QD Josephson junctions presents a greater challenge which has been tackled in a number of recent studies\cite{Paper:VanDam-NAT-2006,Paper:Cleuziou-NATN-2006,Paper:Eichler-PRB-2009,Paper:Lindelof-NL-2007}. In weakly coupled devices where the doublet state dominates the reversal of the supercurrent or $\pi$-junction has been demonstrated\cite{Paper:VanDam-NAT-2006,Paper:Cleuziou-NATN-2006,Paper:Lindelof-NL-2007} in good agreement with theoretical predictions\cite{Paper:Spivak-PRB-2000,Paper:Yeyati-Rodero-2003}. In the strongly coupled Kondo regime the supercurrent has been analyzed through current-biased V-I characteristics in QD Josephson junctions\cite{Paper:Eichler-PRB-2009} and discussed in terms of the zero-bias peak in differential conductance measurements\cite{Paper:Kasper-NJP-2007} with evidence of enhanced critical currents when $t_{K}>1$. In this report we study both the dissipative MAR transport and non-dissipative supercurrents, using a side-gate effect to electrically tune the scaling parameter $t_{K}$.


\begin{figure}[t!]
\includegraphics[width=0.4\textwidth]{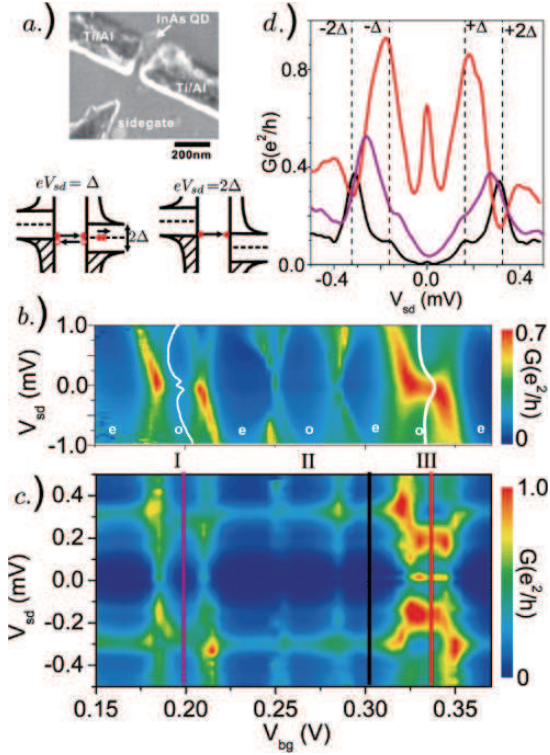}
\caption{(Color online)(\emph{a}) Scanning electron microscope image of the sample studied and energy schematics for the $1^{st}$-order MAR ($eV_{sd}=\Delta$) and single quasi-particle tunneling resonances ($eV_{sd}=2\Delta$). (\emph{b})  Normal state stability diagram ($B=200\,$mT, $V_{sg}=0\,$V). Even (e) and odd (o) electron occupations are indicated. White traces show the Kondo zero bias anomaly in the center of regions I and III. (\emph{c}) Superconducting state stability diagram ($B=0\,$mT, $V_{sg}=0\,$V). (\emph{d}) Superconducting state $G(V_{sd})$ traces for even electron occupation Coulomb blockade ($V_{bg}=0.3\,V$, black trace), the center of region III ($V_{bg}=0.335\,$V, red trace) and the center of region I ($V_{bg}=0.2\,$V, blue trace).}\label{fig:1}
\end{figure}

Devices were fabricated with a single uncapped InAs self-assembled QD with diameter and height of $\sim 100\,$nm and $\sim 20\,$nm, respectively. Conventional \emph{e}-beam lithography and \emph{e}-beam evaporation techniques were used to deposit two Titanium/Aluminium ($5/100\,$nm) electrodes with a nanogap separation of less than $30\,$nm (Fig. \ref{fig:1} (\emph{a})). Additional devices fabrication detail can be found in references \cite{Paper:Buizert1,Paper:Jung2}. We evaluate that $\Delta\sim 162\,\mu$eV and $T_{c}\sim 1.1\,$K (Fig. \ref{fig:1} (\emph{a}) and (\emph{d})). The back-gate is a degenerately Si doped GaAs layer buried $300\,$nm below the sample surface. The side-gate is placed about $200\,$nm away from the QD laterally. Transport measurements were performed in a $^{3}$He-$^{4}$He dilution refrigerator with base temperature $\sim 30\,$mK. The differential conductance was measured using conventional lock-in techniques with an ac excitation of $V_{ac}\sim 3\,\mu$V. For measurement of the supercurrent a four-terminal setup with current-bias was applied.

\begin{figure}[t!]
\includegraphics[width=0.48\textwidth]{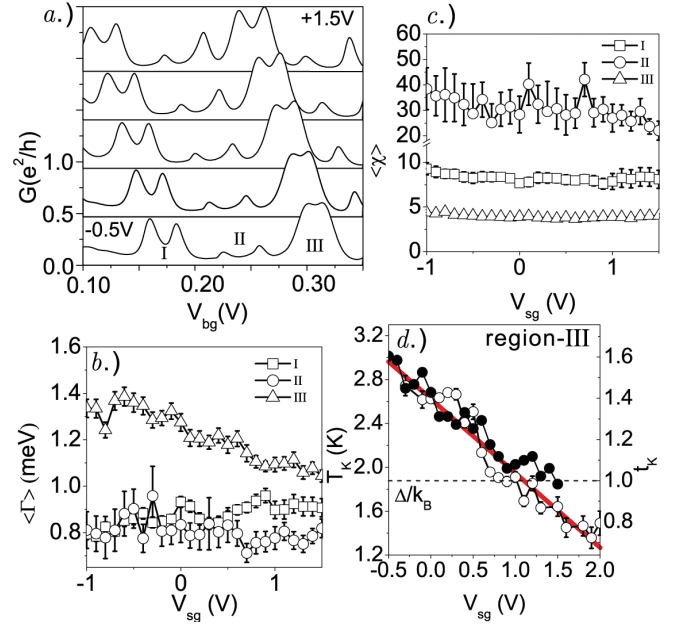}
\caption{(\emph{a}) Normal state Coulomb peaks at $V_{sd}=0\,$V for a range of applied $V_{sg}$. Curves from top to bottom are $V_{sg}=+1.5$ to $-0.5\,$V in $-0.5\,$V steps. Measurements are offset for clarity. (\emph{b}) Summary of normal state $\langle\Gamma\rangle$ for regions I, II and III. (\emph{c}) Summary $\langle\chi\rangle$ for regions I, II and III. (\emph{d}) $T_{K}$ evaluated in region III from the FWHM of the Kondo feature at $B=200\,$mT ($\circ$) and calculated from $\langle\Gamma\rangle$ and $U\sim 2\,$meV ($\bullet$). The horizontal dashed line indicates $\Delta/k_{B}$. The solid line indicates a linear best fit to the FWHM data.}\label{fig:2}
\end{figure}

Fig. \ref{fig:1} (\emph{b}) shows a false colorplot of the differential conductance ($G=dI/dV_{sd}$) taken by sweeping the back-gate $V_{bg}$ and source-drain voltage $V_{sd}$ under an applied magnetic field of $200\,$mT and zero side-gate bias ($V_{sg}=0\,$V). The applied field exceeds the Al lead critical field ($B_{c}\sim150\,$mT) such that the leads are in the normal state. In the $V_{bg}$ range measured we identify three regimes with odd electron occupation, labeled as I, II, and III. The charging energy $U$ is roughly estimated from the width of the Coulomb diamonds to be $2.1, 2.9$ and $2.0\,$meV in regions I, II and III respectively.  We evaluate the backgate leverarm, which relates $V_{bg}$ to the energy in the QD\cite{Paper:Buizert1}, to be $\alpha_{bg}\sim 0.09\,$eV/V. Regions I and III exhibit Kondo zero-bias anomalies which are split and broadened  by the Zeeman energy respectively ($g^{\ast}=6.1\pm 0.2$ and $5.1\pm 0.3$ for regions I and III respectively). In region III we estimate $T_{K}$ from the full width at half maximum (FWHM$=2k_{B}T_{K}$) of Lorentzian fits, with a linear background subtraction, to the Kondo feature\cite{Paper:Goldhaber-NAT-1998}. By subtracting the Zeeman energy from the FWHM a nominal Kondo temperature of $T_{K}=2.6\pm 0.3\,$K ($t_{K}\sim 1.4$) is evaluated. In region I we observe the Kondo feature is already split at $B=200\,$mT indicating the Zeeman energy exceeds $T_{K}$, giving an upper limit of $T_{K}<0.82\pm 0.03\,$K ($t_{K}<0.44$). No Kondo feature is observed in region II indicating that $T_{K}$ is much lower than the measurement temperature.

When $B=0\,$mT, the leads are in the superconducting state, Fig. \ref{fig:1} (\emph{c}). In the even electron occupation regime where Coulomb blockade dominates (Fig. \ref{fig:1} (\emph{d})) prominent resonances at $|eV_{sd}|=2\Delta$ are attributed to direct quasi-particle tunneling between the high density of states at the edge of the superconducting gap in the two leads. A weaker feature at $|eV_{sd}|=\Delta$ is attributed to resonant single Andreev reflections or the $1^{st}$-order MAR resonance (see Fig. \ref{fig:1} (\emph{a})). In odd electron occupation regions the spectrum of MAR features may be renormalized by the Kondo effect resulting in enhanced $1^{st}$-order MAR features at the expense of single quasi-particle tunneling\cite{Paper:Lindelof-PRL-2007, Paper:Schonenberger-PRL-2007,Paper:Buizert1}. Of the regions considered only region III displays enhancement of $1^{st}$-order MAR relative to the $2\Delta$ features, likely due to the higher $T_{K}$. A zero-bias conductance peak is observed in both even occupation regions and regions II and III, however this feature is absent in region I. In regions II and III we can eliminate the Kondo zero-bias anomaly as an origin of the superconducting state zero-bias peak because the magnetic field dependence does not show Zeeman splitting. The zero-bias peak can therefore be regarded as a signature of supercurrent through the device\cite{Paper:Kasper-NJP-2007}. A significant supercurrent feature is observed only in region III providing evidence of enhancement for high normal state $T_{K}$\cite{Paper:Matveev-JETP-1989}.

We will now focus on the effect of the side-gate on the normal state transport. Coulomb oscillations in the normal state for a range of $V_{sg}$ are shown in Fig. \ref{fig:2} (\emph{a}). Coulomb peaks are shifted towards lower $V_{bg}$ as $V_{sg}$ is increased with an evaluated leverarm of $\alpha_{sg} \sim 0.025\alpha_{bg}$. We observe that $V_{sg}$ also alters the lineshape of the Coulomb peaks. Estimates of both the asymmetry in source (S) and drain (D) lead tunnel couplings ($\chi=\Gamma_{S,D}/\Gamma_{D,S}$) and the total tunnel coupling ($\Gamma=\Gamma_{S}+\Gamma_{D}$) are obtained by fitting the even valley part of the Coulomb oscillation peaks using a Lorentzian expression following the method described in reference \cite{Paper:Kasper-NJP-2007} (see supporting information\cite{EPAPS-SQDS}). Results of the average tunnel coupling $\langle\Gamma\rangle$ and average tunnel coupling assymmetry $\langle\chi\rangle$ evaluated for the two Coulomb peaks in all three regions are plotted in Fig. \ref{fig:2} (\emph{b}) and (\emph{c}) respectively. Values of $\langle\chi\rangle$ for regions I and III are found to be fairly constant while region II displays a decrease in $\langle\chi\rangle$ with increasing $V_{sg}$ which accounts for the increase in peak conductance of the fourth Coulomb peak observed in Fig. \ref{fig:2} (\emph{a}). In all regions we observe no significant change in the $g$-factor or $U$ when $V_{sg}$ is altered. We estimate $T_{K}$ in experiment from the FWHM of the Kondo zero-bias anomaly\cite{Paper:Buizert1} in region III for a range of $V_{sg}$ in Fig. \ref{fig:2} (\emph{d}). The dependence of $T_{K}$ on $\Gamma$ in the center of the odd electron occupation region is well formulated as $T_{K}\sim\half\sqrt{U\Gamma}\exp(-\pi U/4\Gamma)$\cite{Paper:Haldane-PRL-1978}. The observed decrease in $T_{K}$ calculated from the FWHM with increasing $V_{sg}$ is consistent with the decrease in that calculated from $\langle\Gamma\rangle$ evaluated from the Coulomb peaks in region III. These results indicate that in region III $T_{K}$ may be tuned by controlling the tunnel coupling via the side-gate. In regions I and II $\langle\Gamma\rangle$ (and $U$) remains relatively constant when $V_{sg}$ is applied and we therefore observed no noticeable effect on $T_{K}$. $\Gamma$ is determined by the effective overlap between lead states and the wavefunction of the confined electrons\cite{Paper:Jung2}. We expect that the side-gate modulates the lateral confinement and displaces the electron wave function, resulting in a change in $\Gamma$. The influence of the side-gate is therefore highly dependent on specific symmetry of the orbital state and the corresponding expansion of the wavefunction, which accounts for the different behavior in regions I, II and III. We conclude that in region III the parameter $t_{K}$ may be tuned in the range $t_{K}\sim 1.5\rightarrow 0.8$.


\begin{figure}[t!]
\includegraphics[width=0.5\textwidth]{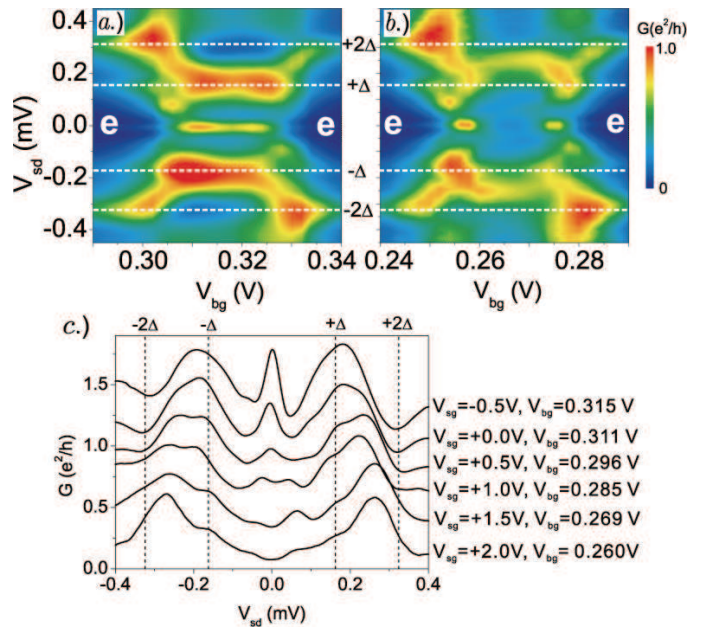}
\caption{(Color online) Plots of $G(V_{sd},V_{bg})$ displaying the subgap transport resonances in region III for $V_{sg}=-0.5\,$V (\emph{a}) and $V_{sg}=+1.5\,$V  (\emph{b}). Horizontal lines indicate bias for quasi-particle tunneling resonance $|eV_{sd}|=2\Delta$ and $1^{st}$-order MAR $|eV_{sd}|=\Delta$. Even (e) occupation regions are identified. (\emph{c}) Plot of $G(V_{sd})$ in the center of region III for a range of $V_{sg}$. Traces from bottom to top are offset by $+0.2\,e^{2}/h$ for clarity.}\label{fig:3}
\end{figure}

We now focus on the superconducting transport in region III as $t_{K}$ is tuned through unity. Details of the effect of $V_{sg}$ in regions I and II are included in the supplemental information\cite{EPAPS-SQDS}. Fig. \ref{fig:3} (\emph{a}) shows a false color plot of differential conductance around region III for $V_{sg}=-0.5\,$V (\emph{a}) and $+1.5\,$V (\emph{b}), focusing on the subgap transport resonances described earlier. While the even occupation regions display similar subgap transport resonances when $V_{sg}=+1.5$ and $V_{sg}=-0.5 V$, the odd occupation region displays pronounced differences. The zero-bias conductance peak is observed for $V_{sg}=-0.5\,$V but not for $V_{sg}=+1.5\,$V, indicating a suppression of the supercurrent when $V_{sg}$ is increased. Fig. \ref{fig:3} (\emph{c}) displays plots of $G(V_{sd})$ at the center of the region III for a range of $V_{sg}$. The magnitude of the zero-bias peak is gradually reduced and disappears at around $V_{sg}\sim +0.8\,$V. When the zero-bias peak diminishes a pair of small resonances are observed near $V_{sd}=0$. These resonances may be attributed to higher order MAR. We also observe a distinct shift in the most prominent transport resonance from the $1^{st}$-order MAR feature at $|eV_{sd}|=\Delta$ to an intermediate feature between $|eV_{sd}|=2\Delta$ and $|eV_{sd}|=\Delta$. Note also that the minima at $|eV_{sd}|=2\Delta$ are suppressed as $V_{sg}$ is increased. Viewed in the context of recent studies of the interplay between MAR and the Kondo effect\cite{Paper:Buizert1,Paper:Schonenberger-PRL-2007,Paper:Lindelof-PRL-2007} we attribute the pronounced change in the resonances to a reduction of the influence of Kondo effect on the spectrum of MAR resonances. We note similarity between the subgap resonances in region III for high $V_{sg}$ (indicating low $T_{K}$) with those observed in region I (shown in Fig. \ref{fig:1} (\emph{d})) where $t_{K}<0.44$.

\begin{figure}[t!]
\includegraphics[width=0.5\textwidth]{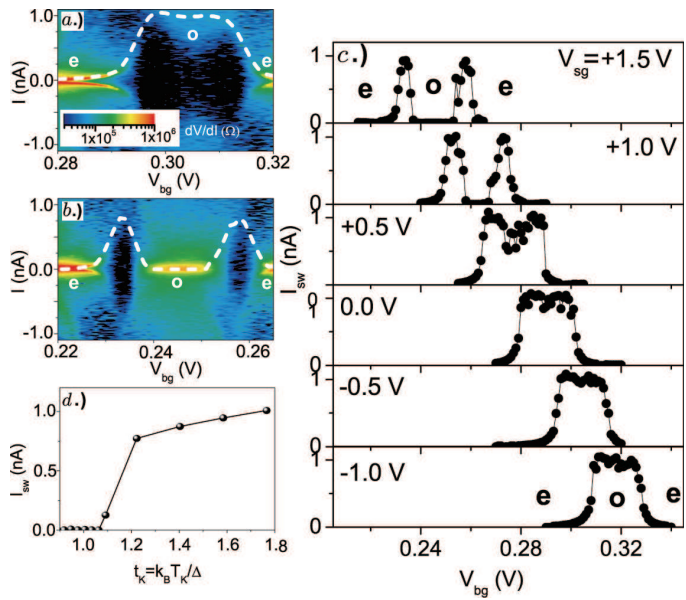}
\caption{(Color online) Plot of $dV/dI$ for four terminal $I$-bias measurements in region III with $V_{sg}=-0.5\,$V (\emph{a}) and $V_{sg}=+1.5\,$V (\emph{b}). Dashed white lines indicate the trend of $I_{sw}$. (\emph{c}) $I_{sw}(V_{bg})$ in region III for a range of $V_{sg}$. (\emph{d}) $I_{sw}(t_{K})$ for the center of region III. $t_{K}$ is estimated from the linear fit in Fig. \ref{fig:2} (\emph{d}). Even (e) and odd (o) occupation is indicated.}\label{fig:4}
\end{figure}

We analyze the transport further using a four terminal current bias measurement to study the supercurrent. We plot the four-terminal differential resistance ($dV/dI$) as a function of $V_{bg}$ and current ($I$) for $V_{sg}=-0.5\,$V and $V_{sg}=+1.5\,$V in Fig. \ref{fig:4} (\emph{a}) and (\emph{b}) respectively. When measuring the non-dissipative current in the junction we must consider the dissipation in the surrounding circuit\cite{Book:Tinkham} which determines the characteristics of the V-I curves. From consideration of the RCSJ model\cite{Paper:Jarillo-Herrero-NAT-2006,Paper:GangLiu-PRL-2009,Paper:Lindelof-NL-2007,Paper:Eichler-PRB-2009} we determine that in region III the junction is heavily overdamped\cite{EPAPS-SQDS} with a non-zero resistance due to thermal phase diffusion\cite{Paper:Ingold-PRB-1994}. We qualitatively evaluate the relative evolution of the junctions intrinsic critical current using a 'switching current' ($I_{sw}$), taken as the current value at the maximum differential resistance (white dashed lines in Fig. \ref{fig:2} (\emph{a} and {b})).  $I_{sw}$ is influenced by the dissipation of the extrinsic circuit and is expected to be significantly reduced from the intrinsic critical current due to thermal fluctuations. In Fig. \ref{fig:4} (\emph{a}) where $V_{sg}=-0.5\,$V, $I_{sw}$ is high in the odd occupation region and small but non-zero in the even occupation regions where Coulomb blockade dominates. Fig. \ref{fig:4} (\emph{c}) shows the evolution of $I_{sw}(V_{bg})$ for a range of $V_{sg}$. In good agreement with the $V$-bias measurements we find that supercurrent feature is reduced to zero in the odd electron occupation region when $V_{sg}$ is high. Recent functional\cite{Paper:Karrasch-PRB-2008} and numerical\cite{Paper:Choi-PRB-2004,Paper:Oguri-JPSJ-1994} renormalization group studies predict a sharp drop in absolute $I_{c}$ when the device is tuned to the magnetic doublet regime. The overall trend of $I_{sw}(V_{bg})$ on $V_{sg}$ shown in Fig. \ref{fig:4} (\emph{d}) therefore qualitatively matches the scenario of a phase transition between a Kondo dominated singlet state to a degenerate 'magnetic' doublet. In the magnetic doublet regime ($t_{K}<1$) the supercurrent is strongly suppressed by Coulomb blockade while in the Kondo singlet regime ($t_{K}>1$) the supercurrent may be enhanced by the Kondo effect. The transition is observed at $t_{K}\sim 1.1$ in good agreement with the disappearance of the zero-bias peak in voltage bias measurements. Choi \etal\cite{Paper:Choi-PRB-2004} used the numerical renormalization group to calculate $I_{c}$ as a function of $t_{K}$ and predicted the transition at $t_{K}\sim 2$. For $t_{K}>2$ a saturation of $I_{c}$ was predicted. Siano and Egger\cite{Paper:Siano-PRL-2004} applied the Hirsch-Fye Monte Carlo method to predict a transition at $t_{K}\sim 1.1$. Our experimental result is in good qualitative agreement with the predicted features in both of these studies. Discrepancies may arise from the non-ideal nature of real devices in which tunnel coupling is assymmetric and extrinsic environmental effects may dominate. In some measurements\cite{Paper:Lindelof-NL-2007} the low supercurrent in the magnetic doublet ($\pi$-junction) regime (when $t_{k}\ll 1$) has been measured, however in the device considered here no supercurrent branch is observed implying that coherent processes are overcome by dissipation in the circuit and thermal fluctuations due to a poorly screened electromagnetic environment.




In the present report the side-gate electrode is demonstrated to allow limited tuning of the device parameters in a region of constant electron number. This technique is used to elucidate the effect of Kondo correlations on both the supercurrent and subgap dissipative transport around the phase transition between magnetic double and Kondo singlet states. The onset of the transition was observed at $t_{K}=k_{B}T_{K}/\Delta\sim 1.1$.

\begin{acknowledgements}
We acknowledge valuable discussions with A. Oguri and Y. Tanaka. We acknowledge financial support from the Japan Society for the Promotion of Science, grant XXXX (R.S.D.), Grant-in-Aid for Scientific Research S(No. 19104007) and A(No. 21244046) and QuEST program (BAR-0824) (S.T.).
\end{acknowledgements}


\end{document}